\documentclass{appolb}
\usepackage{epsfig}
\usepackage{color}
\usepackage{cite}
\usepackage{graphicx}
\usepackage{subcaption}
\usepackage{amsfonts}
\usepackage{rotating}
\begin{document}
\pagestyle{plain}

\newcommand{\var}{\mathsf{var}}
\newcommand{\cov}{\mathsf{cov}}
\newcommand{\vc}{\mathbf}
\newcommand{\bm}{\mathbf}
\newcommand{\E}{\mathsf{E}}

\newcount\eLiNe\eLiNe=\inputlineno\advance\eLiNe by -1
\title {Key courses of academic curriculum uncovered by data mining of students' grades}
\author{\L. G. Gajewski, J. Cho\l oniewski\footnote{e-mail: choloniewski@if.pw.edu.pl} 
\address{Faculty of Physics, Center of Excellence for Complex Systems Research,\\Warsaw University 
of Technology \\ 
Koszykowa 75, PL-00662 Warsaw, Poland}
\\ \vspace{10pt}
{
J. A. Ho\l yst
\address{Faculty of Physics, Center of Excellence for Complex Systems Research,\\Warsaw University of Technology\\ Koszykowa 75, PL-00662 Warszawa, Poland}
\address{ITMO University, \\19, Kronverkskiy av., 197101 Saint Petersburg, Russia}
}}
\maketitle

\begin{abstract}
Learning is a complex cognitive process that depends not only on an individual capability of knowledge absorption but it can be also influenced by various group interactions and by the structure of an academic curriculum. We have applied methods of statistical analyses and data mining (Principal Component Analysis and Maximal Spanning Tree) for anonymized students' scores at Faculty of Physics, Warsaw University of Technology. A slight negative linear correlation exists between mean and variance of course grades, i.e. courses with higher mean scores tend to possess a lower scores variance.
 There are courses playing a central role, e.g. their scores are highly correlated to other scores and they are in the centre of corresponding Maximal Spanning Trees. Other courses contribute significantly to students' score variance as well to the first principal component and they are responsible for differentiation of students' scores. Correlations of the first principal component to courses' mean scores and scores variance suggest that this component can be used for assigning ECTS points to a given course. The analyse is independent from declared curricula of considered courses. The proposed methodology is universal and can be applied for analysis of student's scores and academic  curriculum at any faculty.
\end{abstract}
\PACS{89.65.-s, 89.65.Ef}
\section{Introduction}

Teaching and learning play a pivotal role for development of an individual as well as of the human civilization\cite{history}. In fact, we spend about 18 years for our education and complexity of this process can be studied at different levels, e.g. scores of an individual pupil/student, scores of learning groups, scores related to different courses, changes of scores during the study periods but also early drop-out prediction\cite{dropout} etc. \cite{edm}. Advancements in various platforms for e-learning (such as Massive Open On-line Courses \cite{moocs}) make some analyses obvious, but a progress in digital tools related to educational processes (such as USOS\cite{usos} -- virtual deanery) and their growing popularity make possible new kinds of analyses where data mining methods can be applied. Results of such analyses can lead to improvement of teaching curricula and study plans at schools/universities. 

In the present paper, we analyse grades at Faculty of Physics, Warsaw University of Technology during the first five semesters of the first level studies leading to the Engineering Degree in Applied Physics. The core of curriculum consists mostly of courses in physics, mathematics, computer science and electrical engineering. After the 5th semester a student selects one of specializations: Optoelectronics, Materials and Nanostructures (formerly Solid State Physics), Computer Physics, or Medical Physics. We are interested to learn dependencies between students' scores at different courses. To uncover these dependencies we shall use specific tools of data mining methods which are widely known and applied in complex systems analyses, namely correlation matrices\cite{corrmat,speth}, correlation--based networks \cite{dalmaso}, minimal (maximal) spanning trees \cite{tuminello,onnela,gorski}, and principal component analysis \cite{giuliani,suslick,disney,power}.

As far as we know, this is the first study of teaching results at Polish universities using data mining tools. A similar methodology was applied for learning data at Singaporean primary schools \cite{singapore} but it was focused on dynamics of single lessons rather than on a whole teaching  curriculum. We hope that the framework  proposed in this paper can be applied to data from other faculties and can be helpful to better understand and organize teaching/learning process.

\section{Data acquisition and filtering}
The data set has been received from the deanery of Faculty of Physics at Warsaw University of Technology and it has contained information about students' final grades of particular courses as well as students' group and specialization. The data contains records of engineering degree candidates since 2010 until 2015.

We have excluded from the analysis all students who did not pass the first five semesters and we have chosen three consecutive students' classes: J, K and L. The class J means all students that started their academic studies in 2010, K in 2011 and L in 2012. Those classes have been chosen because only for them there was a large number of records of core courses that every student must complete in order to get a degree. After removing general University courses like physical education or foreign languages we have got $27$ core courses (see Table \ref{tab:corecourses}). Each student (and course) in our analysis is represented as a vector of numbers –- grades. The total number of students considered in the analyses is 217 (class J -- 76 students, K -- 66, L -- 75).

Since there have been still some missing records for individual students, we have filled those gaps with an average score at that course in all three classes. Amounts of missing grades at each class are: class J -- 5.46\%, K -- 14.03\%, L -- 1.68\%, and for all classes combined -- 6.72\%.

Abbreviations of courses names as well as their mean grades and variances of grades (and other calculated parameters) are listed in Table~\ref{tab:corecourses}.

\begin{sidewaystable}
\resizebox{\columnwidth}{!}{%
\begin{tabular}{l*{2}{c}{c}{c}{c}{c}{c}{c}{c}r}
Course & Abbreviation & Semester & Mean Grade & Variance & Sum of Correlations & Explained Variance & 1st PCC & Node Degree (MST) \\
\hline
Algebra and Geometry & AG & 1 & 3.72 & 0.45 & 10.93 & 0.033 & 0.21 & 2\\
Analysis of Experimental Data & AED & 1 & 4.12 & 0.25 & 8.93 & 0.018 & 0.12 & 1\\
Fundamentals of Information Technology & FIT & 1 & \textbf{4.58} & \underline{0.18} & 7.45 & \underline{0.013} & \underline{0.090} & 1\\
Fundamentals of Physics 1 & FP1 & 1 & 3.51 & 0.48 & \textbf{12.29} & 0.035 & 0.24 & \textbf{9}\\
Mathematical Analysis 1 & MA1 & 1 & 3.49 & 0.64 & 10.18 & 0.046 & 0.25 & 1\\ \hline
Basics of Programming & BP & 2 & 4.18 & 0.53 & 8.06 & 0.038 & 0.18 & 1\\
Fundamentals of Physics 2 & FP2 & 2 & 3.62 & 0.51 & \textbf{11.52} & 0.037 & 0.24 & 3\\
Mathematical Analysis 2 & MA2 & 2 & \underline{3.45} & 0.73 & 11.2 & 0.053 & \textbf{0.29} & 3\\
Physics Laboratory 1 & PL1 & 2 & 4.07 & 0.3 & 7.92 & 0.021 & 0.12 & 1\\ \hline
Fundamentals of Electronics & FE & 3 & 3.56 & 0.67 & 10.1 & 0.048 & 0.24 & 1\\
Mathematical Analysis 3 & MA3 & 3 & \underline{3.43} & 0.79 & 10.03 & 0.057 & 0.27 & 1\\
Mechanics & M & 3 & 3.85 & 0.49 & 10.18 & 0.035 & 0.21 & 3\\
Programming Languages & PL & 3 & 4.03 & \textbf{0.89} & 8.31 & \textbf{0.064} & 0.24 & 3\\ \hline
Electrodynamics & E & 4 & 4.37 & 0.35 & 8.03 & 0.025 & 0.14 & 1\\
Electronics in Physical Experiment & EPE & 4 & \textbf{4.55} & \underline{0.19} & 7.09 & \underline{0.014} & \underline{0.077} & 2\\
Mathematical Methods of Physics & MMP & 4 & 4.16 & 0.54 & 8.68 & 0.039 & 0.18 & 1\\
Object-Oriented Programming & OOP & 4 & 4.21 & 0.47 & 8.06 & 0.034 & 0.15 & 1\\
Probability & P & 4 & 3.56 & 0.63 & 9.25 & 0.046 & 0.23 & 1\\
Quantum Physics & QP & 4 & 3.70 & \textbf{0.97} & 9.64 & \textbf{0.070} & \textbf{0.29} & 2\\ \hline
Chemistry & C & 5 & 4.02 & 0.38 & 9.55 & 0.027 & 0.16 & 2\\
Engineering Graphics & EG & 5 & 3.72 & 0.51 & \underline{5.97} & 0.036 & 0.11 & 1\\
Fundamentals of Optics & FO & 5 & 3.62 & 0.5 & \underline{5.01} & 0.036 & 0.091 & 1\\
Fundamentals of Virtual Devices Design & VDD & 5 & 3.48 & 0.6 & 8.58 & 0.044 & 0.20 & 2\\
Introduction to Solid State Physics & ISSP & 5 & 3.97 & 0.6 & 7.18 & 0.044 & 0.18 & 1\\
Introduction to Nuclear Physics & INP & 5 & 3.92 & 0.28 & 8.03 & 0.020 & 0.11 & 2\\
Physics Laboratory 2 & PL2 & 5 & 4.24 & 0.27 & 7.54 & 0.020 & 0.11 & 2\\
Statistical Physics and Thermodynamics & SPT & 5 & 4.09 & 0.67 & 8.21 & 0.048 & 0.19 & 3\\ \hline
\end{tabular}
}
\caption{Core courses chosen for analysis -- names, abbreviations used in figures, semester at which they take place, various parameters calculated in the paper. Two highest values in columns 4-8 are \textbf{bold}, and the two lowest are \underline{underlined}. In the last column, only the highest value has been highlighted. \textit{1st PCC} stands for "1st Principal Component contribution".}
\label{tab:corecourses}
\end{sidewaystable}

\section{Methodology}
As a number of core courses is the same for each student, the dataset can be presented as $N_S\times N_C$ matrix of grades $G_{s,c}$ (where $N_S$ -- number of students, $N_C$ -- number of courses, $s$ -- student index, $c$ -- course index). Positions of columns and rows are arbitrary but set. 

Such a representation allows us to easily define a measure of a coincidental deviations from the means between courses as a Pearson product-moment correlation coefficient \cite{hogg}:
\begin{equation}
\label{corr}
 C_{X,Y} = {{\cov(X,Y)} \over {\sigma_X \sigma_Y}}
\end{equation}
where in our case $X=G_{*,X}$ and $Y=G_{*,Y}$ are vectors of grades for given courses $X$ and $Y$ obtained by students which can be perceived as series of observations of random variables, $\mathrm{cov}(X,Y)$ is a covariance between these grades X and Y, and $\sigma_\alpha$ is a standard deviation from the mean of the variable $\alpha$. A positive (negative) correlation $C_{X,Y}$ between two courses is when variations from corresponding means tend to have the same (opposite) sign for a given student. 

A courses' network can be built using calculated values $C_{X,Y}$ as links' weights. Such a  network is unfortunately a hardly readable weighted complete graph. To overcome this problem we have applied the inverted Kruskal's algorithm \cite{kruskal} and we have reduced the complete graph to the related maximal spanning tree (MST). Let us remind that the MST possesses  the same number of nodes as the original network and the inverted Kruskal's algorithm sorts a list of edges from the original network by correlations $C_{X,Y}$ descending and iterates through the list adding an edge if it does not cause a cycle to occur.

The Principal Component Analysis \cite{hastie} is a method of variables transformation equivalent to a special rotation of coordination axes. Let us write  scores of a given student as a vector variable $\vc{x}$ where the dimension of this vector equals to the number of all courses the student attended.
One should start with centring the analysed variable $\vc{x}$:
\begin{equation}
\label{pca_centering}
\tilde{\vc{x}} = \vc{x} - \E(\vc{x})
\end{equation}
Here $\E(...)$ stands for the vector of  expected values, that is calculated as a vector  of  means over all students for each course's score.
The first principal component  is a direction $\gamma_{(1)}$ in the new framework where  the variance of all observed data points projected onto this direction is maximal:
\begin{equation}
\var \left( \bm{\gamma}^{T}_{(1)} \tilde{\vc{x}} \right) = \max_{\vc{a} \in \mathbb{R}^p, \|\vc{a}\|=1} \left\{ \var \left( \vc{a}^{T} \tilde{\vc{x}} \right) \right \},
\end{equation}
where $\gamma_{(1)}^T$ -- the transverse vector defining the first axis of transformed data, $\var(...)$ -- variance of a given variable.
The axis of the second principal  component $\gamma_{(2)}$ is orthogonal to the first one:
\begin{equation}
\bm{\gamma}^{T}_{(1)}  \bm{\gamma}_{(2)}= 0
\end{equation}

and it maximises the remaining part of variance:
\begin{equation}
\var \left( \bm{\gamma}^{T}_{(2)} \tilde{\vc{x}} \right) = \max_{\vc{a} \in \mathbb{R}^p, \|\vc{a}\|=1} \left\{ \var \left( \vc{a}^{T} \tilde{\vc{x}} \right) \right \}
\end{equation}
and so on. In fact the transformation is equivalent to finding eigenvectors of covariance matrix corresponding to maximal eigenvalues (they are real because the matrix is real and symmetric). In such a way one can reduce an original highly dimensional space to a few directions corresponding to largest variances \cite{witten}.

For the sake of comparison, two types of shuffling have been performed. The first type (S -- "by students' grades") is to replace each row of a matrix of grades $G_{s*}$ with its random permutation.
Similarly the second type (C -- "by courses' grades") is to replace each column of the matrix $G_{*c}$ with its random permutation. While the type S shuffling keeps the mean grade of a student the same,  one can expect it destroys possible relations between courses. Exact opposite is for the type C shuffling (keeping courses' means and destroying possible relations between students). 

\section{Results and discussion}
\begin{center}
\begin{figure}[!htb]
\begin{subfigure}[b]{0.31\textwidth}
\includegraphics[height = 4cm]{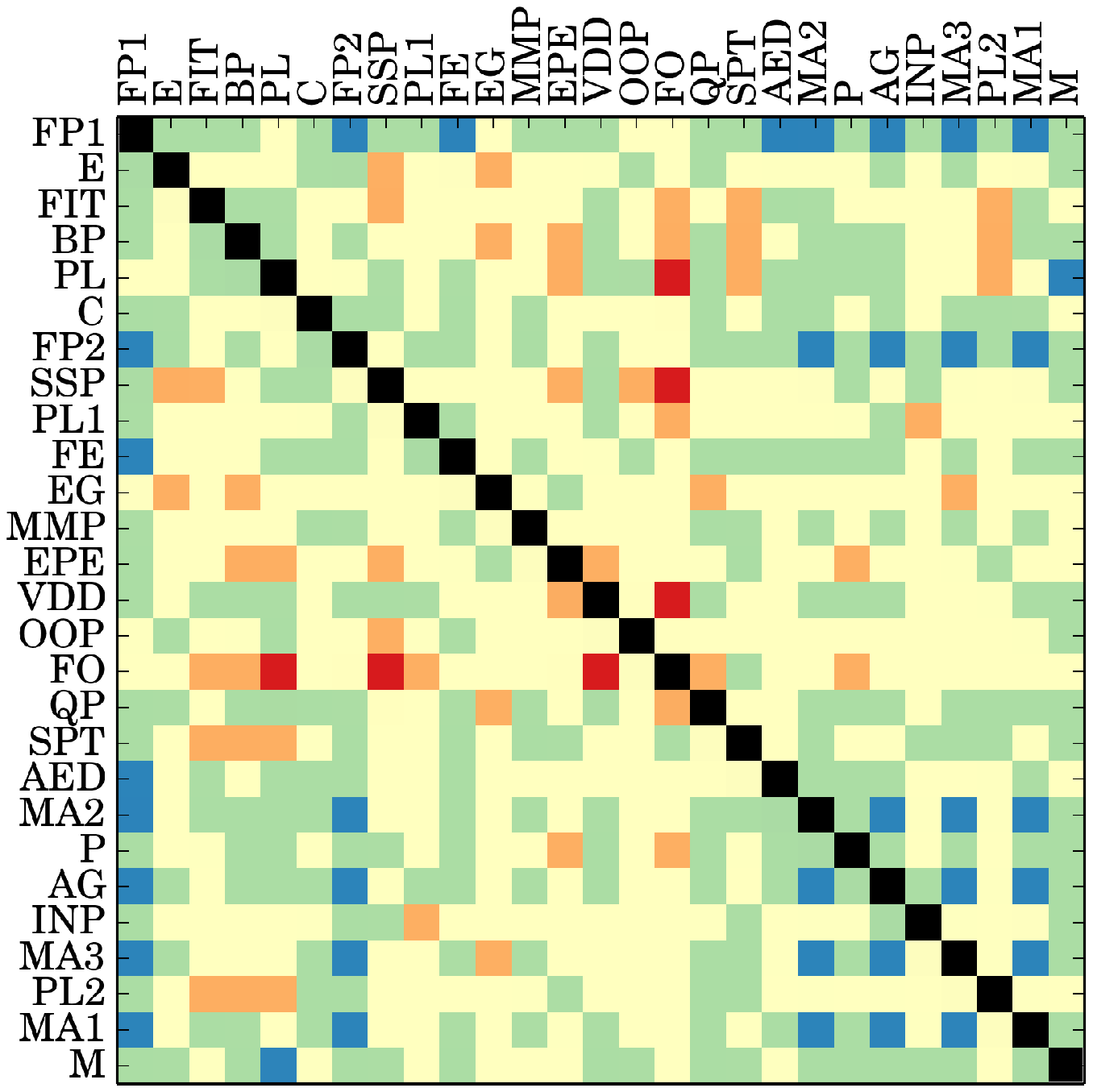}
\caption{original data}
\label{fig:corrr}
\end{subfigure}
\begin{subfigure}[b]{0.28\textwidth}
\includegraphics[height = 4cm]{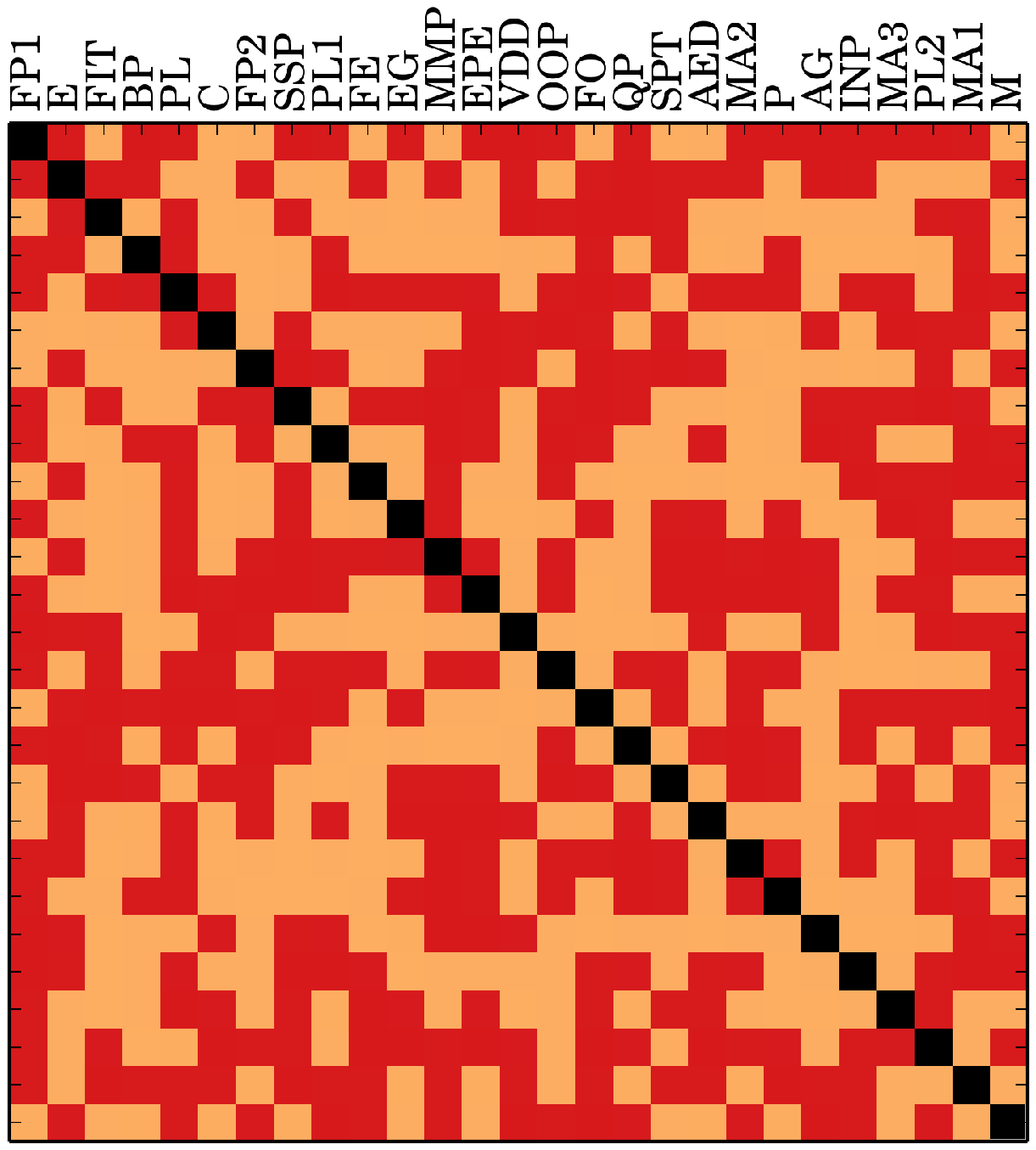}
\caption{C-type shuffling}
\label{fig:corrc}
\end{subfigure}
\begin{subfigure}[b]{0.39\textwidth}
\includegraphics[height = 4cm]{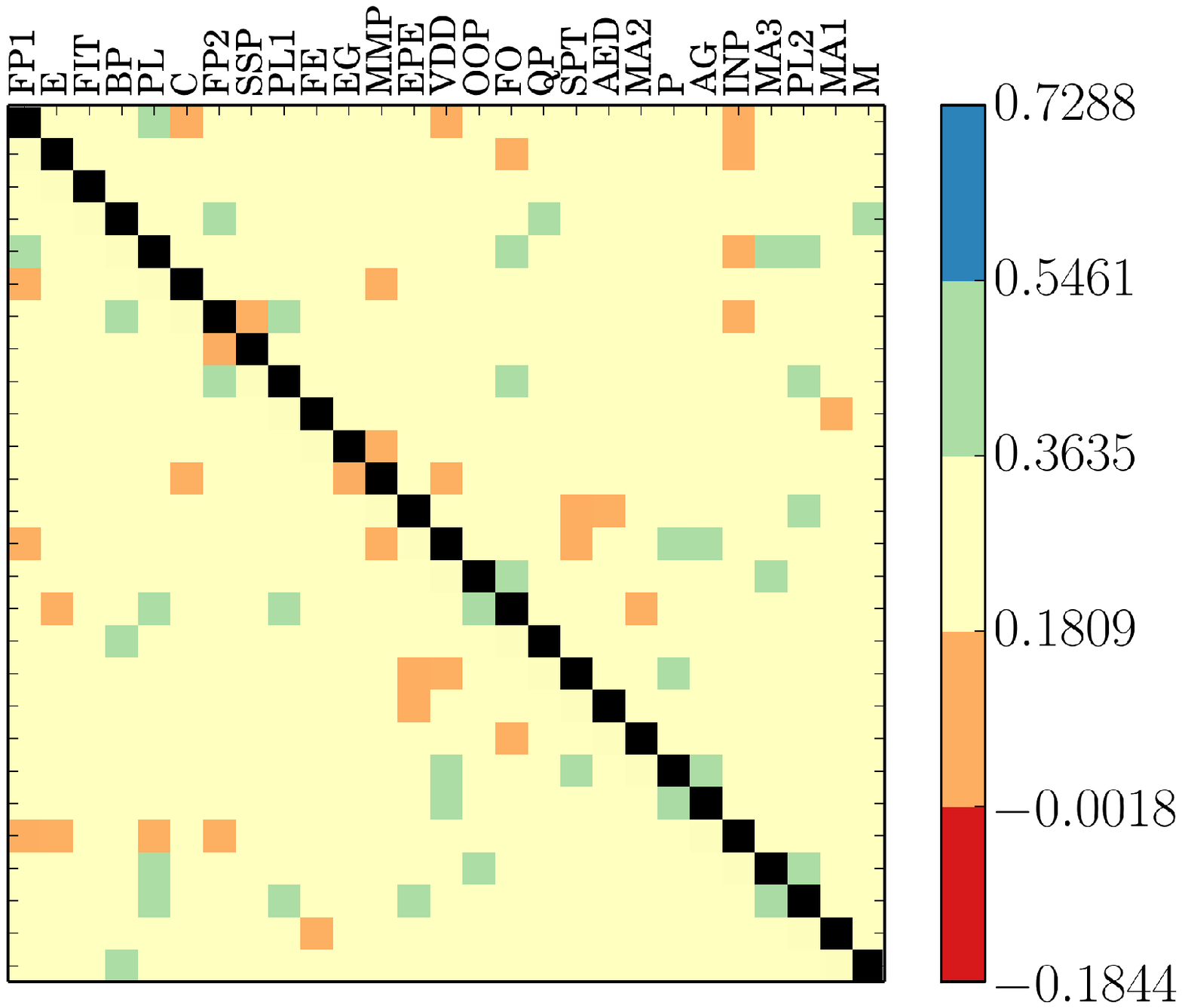}
\caption{S-type shuffling}
\label{fig:corrs}
\end{subfigure}
\caption{Course-course Pearson's correlation coefficients in a form of heatmap. In the C-type shuffling all grades for a given course are randomly exchanged between students and in the S-type shuffling all grades of a given student are randomly exchanged between courses.}
\label{fig:corr}
\end{figure}

\end{center}

Correlation coefficients given by Eq.\ref{corr} have been calculated for each pair of courses and are presented as a heat map in Fig.~\ref{fig:corrr}. The courses' grades are quite well correlated, the mean correlation (excluding diagonal elements) and its standard error is $ 0.326 \pm 0.006$.
There is a triple of courses FP1, FP2 and AG that are strongly correlated one to another, i.e. $C_{FP1,FP2}=0.73$, $C_{FP1,AG}=0.7$ and $C_{FP2,AG}=0.62$. Furthermore courses from this group are the most correlated with other courses (sum of correlation coefficients $C_{X}=\sum\limits_{i\neq X}C_{X,i}$): $C_{FP1}=12.29$, $C_{FP2}=11.52$ and $C_{AG}=10.93$. On the other hand courses FO and EG are least correlated with others; $C_{FO}=5.01$ and $C_{EG}=5.97$ (for all $C_{X}$ -- see Table~\ref{tab:corecourses}). FO is the only course that tends to negatively correlate with some other courses: PL (-0.12), VDD(-0.10) and SSP (-0.06), (for all $C_{X}$ -- see Table~\ref{tab:corecourses}). 

In Figs.~\ref{fig:corrc} and~\ref{fig:corrs} correlation matrices are shown for shuffled data. The effect of C-shuffling is a complete destruction of correlation matrix. New correlations values are close to zero and the mean correlation is $0.0001 \pm 0.007$. 

The S-shuffling diminishes the original correlations by $20\%$, and their mean value is equal to $0.258 \pm 0.005$. It can suggest that a given student gets similar grades for several courses and interchanging his own grades does not change dramatically correlations between grades of different courses. The S-shuffling destroys however high positive correlations of the triple FP1, FP2 and AG as well negative correlations for the course FO. 

In order to further investigate the correlations we have constructed a complete graph of courses where weights of links between nodes are defined by eq.~(\ref{corr}). Corresponding maximal spanning trees (MST) for this graph are presented in Fig. \ref{fig:mstjkl} (merged data from classes J, K, L) and Fig. \ref{fig:mstl} (only students of L - class) 

For better readability labels above links are $C_{X,Y}$ and thickness of a given link is proportional to that value. One can clearly see that the course FP1 is a central hub of the network. Its degree in Fig. \ref{fig:mstjkl} is $k_{JKL}^{MST}=9$ and in Fig. \ref{fig:mstl} $k_{L}^{MST}=6$. One can see also that courses FP2 and AG from the highly correlated triple {FP1, FP2, AG} are in the centres of MST in Figs.\ref{fig:mstjkl} and \ref{fig:mstl}. At the second figure these two courses are local hubs of degrees $4$ and $5$ respectively. We need to stress that observed MSTs of different classes are different and certain courses can appear in different neighbourhood depending on what class we consider. In all cases however the course FP1 remains a central node of MST and the courses FP2 and AG are close to it. It suggests that this triple of courses lays a foundation of knowledge for the students and perhaps in some way it verifies students' knowledge acquired in secondary schools since it is on the first two semesters. We shall call these three courses {\it central} (C). 

\begin{center}
\begin{figure}[!htb]
\includegraphics[width=\textwidth]{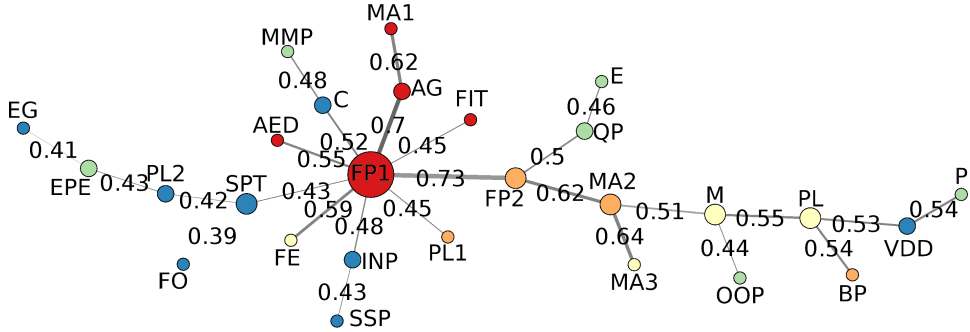}
\caption{Maximal spanning tree for the network of courses where links are defined via correlations between the nodes. Symbols correspond to different courses (see Table \ref{tab:corecourses}), numbers indicate correlations coefficients between courses and  colours are indicators of a corresponding semester for that course (i.e. red - semester 1, blue - semester 5 etc., see Tab. \ref{tab:corecourses}). The network has been constructed for all three classes (J,K,L) combined. Size of a node is proportional to its degree, i.e. to a number of its nearest neighbours in the MST.}
\label{fig:mstjkl}
\end{figure}

\begin{figure}[!htb]
\includegraphics[width=\textwidth]{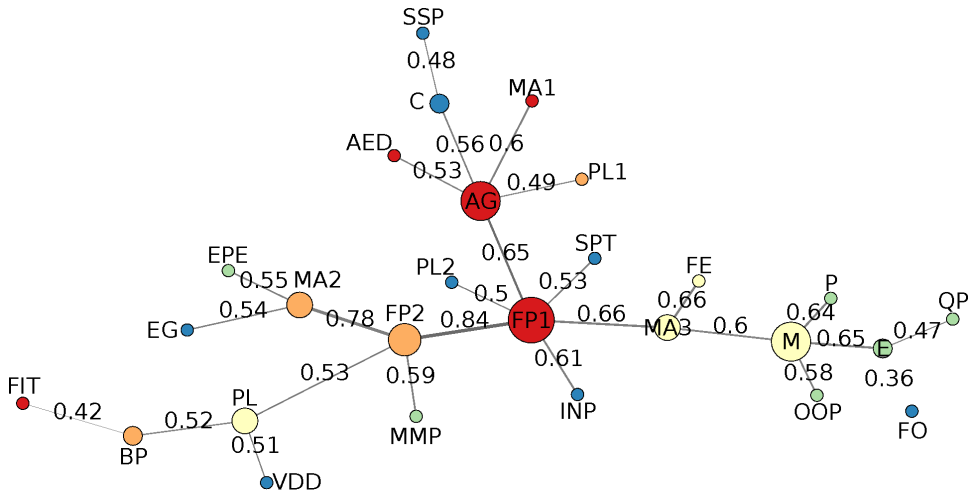}
\caption{Maximal spanning tree for one class (L) only. Meaning of symbols, labels and  colours is the same as for  Fig.~\ref{fig:mstjkl}}
\label{fig:mstl}
\end{figure}
\end{center}

\begin{figure}[t]
\centering
\includegraphics[width=.7\textwidth]{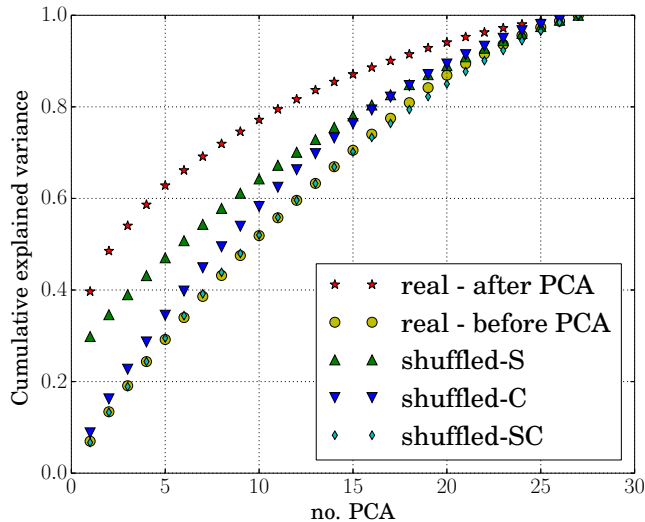}
\caption{Plot of explained cumulative variance.  We accumulate variances of each course  (labelled {\it real-before PCA}) in descending order, normalized in such a way that the total sum is $1.0$. After PCA decomposition we do similarly with principal components (points labelled {\it real-after PCA}). Points labelled {\it shuffled-S/C/SC} are results of PCA decomposition after a particular shuffling. S is for S-type shuffling, C for  C-type shuffling and SC for  both shuffles together.}
\label{fig:pcacum}
\end{figure}

\begin{figure}[t]
\centering
\includegraphics[width=\textwidth]{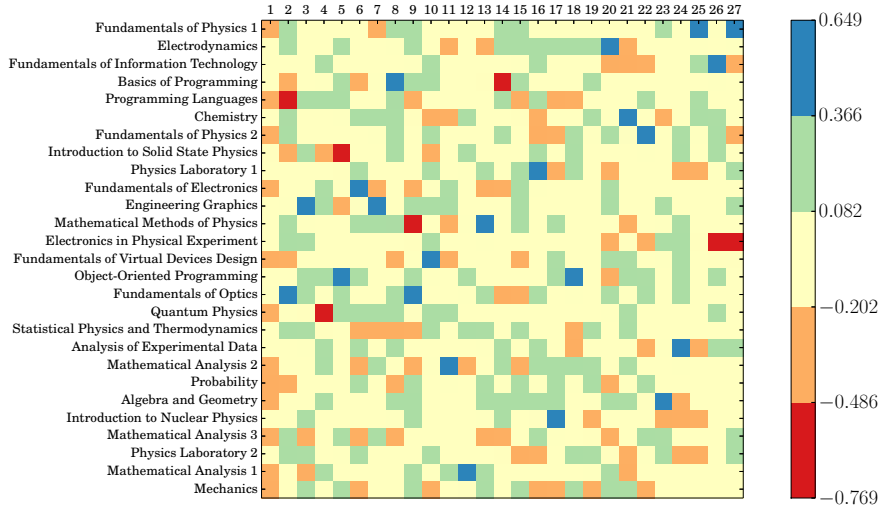}
\caption{Transformation matrix from original variables to principal components. It shows what contribution each original variable (a course) has in a particular principal component.}
\label{fig:pcamat}
\end{figure}

Up to now we have focused on correlations between scores of various courses. Important information is contained also in scores' variances and in spectral properties of entire covariance matrix. Variances of each course have been collected in Table~\ref{tab:corecourses} and  Principal Component Analysis (PCA) has been applied to get insight about a possible dimensionality reduction, i.e. what variables are needed to describe student's scores accurately enough. We have taken grades of students as observations and courses as features (observables). A comparison of variances explained by variables (sorted by its variance descending) for the original data, PCA-transformed original and shuffled data is shown in Fig.~\ref{fig:pcacum}. Matrix in Fig. \ref{fig:pcamat} presents contributions of each course to Principal Components (PC). The First PC explains almost 40\% of total variance (Fig.~\ref{fig:pcacum}) and all courses contributions to it have the same sign (Fig.~\ref{fig:pcamat}). It suggests that PC is connected to a student's mean grade and can be used as a simple discrimination between students. Further principle components explain slightly higher amount of variance than most variate courses (namely QP -- 7\%). The second principal component has unlike contributions of PL and FO what was visible in the correlations analysis as well. The highest contribution to the third PC has been made by EG, to the fourth -- QP and so on. The S-shuffling of students grades diminishes the role of the first PC that explains $30\%$ of the total variance (Fig.~\ref{fig:pcacum}) and the first PC after C-shuffling explains only $10\%$ of the total variance. These facts are in agreement with observations of shuffling influence on Pearson's correlation coefficients (Figs.\ref{fig:corrr}, \ref{fig:corrc}, \ref{fig:corrs}) where the S-shuffling was less destructive than the C-shuffling. Let us remind also that after combining S- and C-shuffling the plot of cumulative variance is indistinguishable from the corresponding plot using variances of original study courses (Fig.~\ref{fig:pcacum}). 

\begin{figure}
\centering
\begin{subfigure}[t]{0.5\textwidth}
\includegraphics[height=4.4cm]{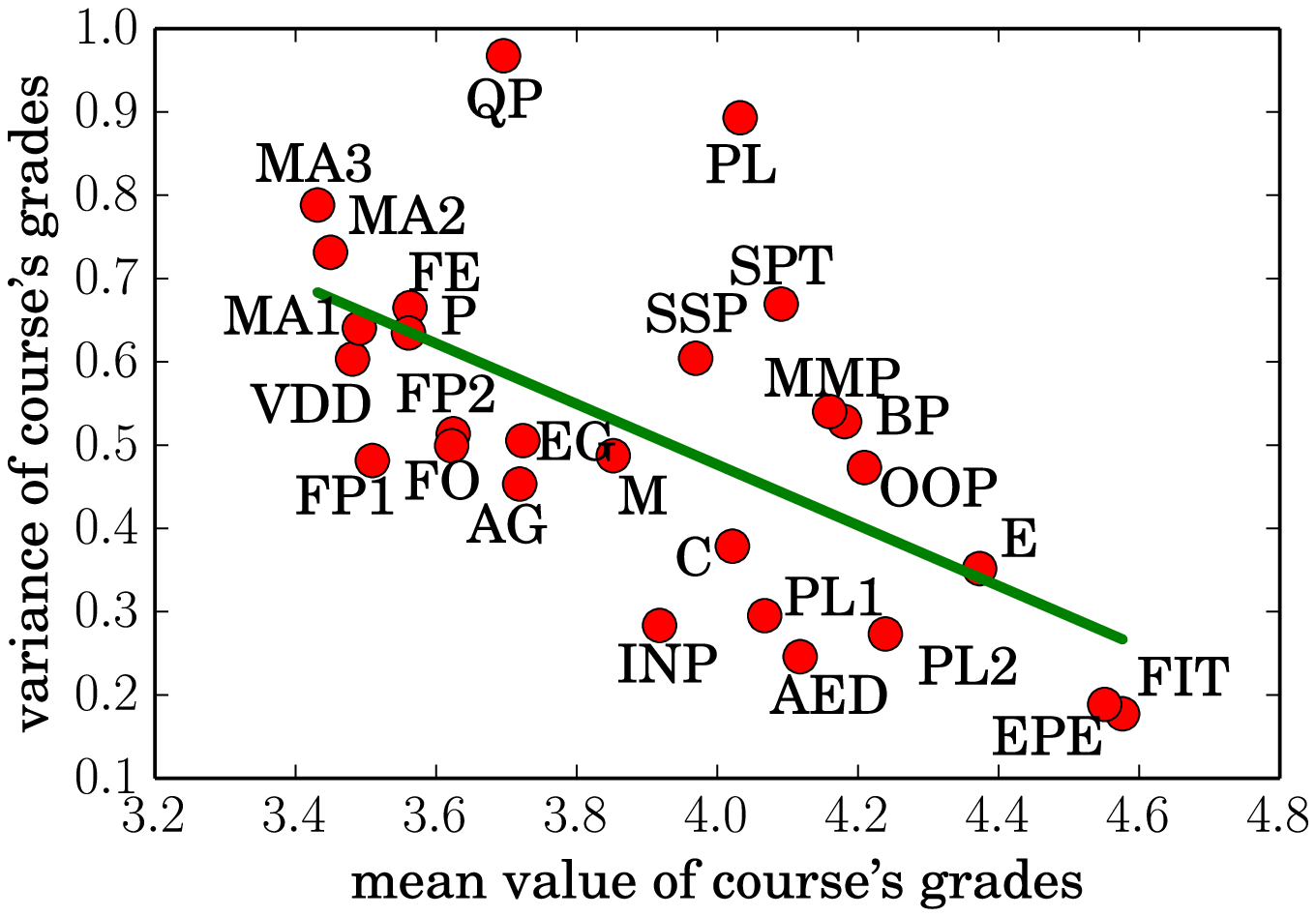}
\caption{$a = -0.36$ $R^2 = 0.37$}
\label{fig:avsvar}
\end{subfigure}
\begin{subfigure}[t]{0.450\textwidth}
\includegraphics[height=4.4cm]{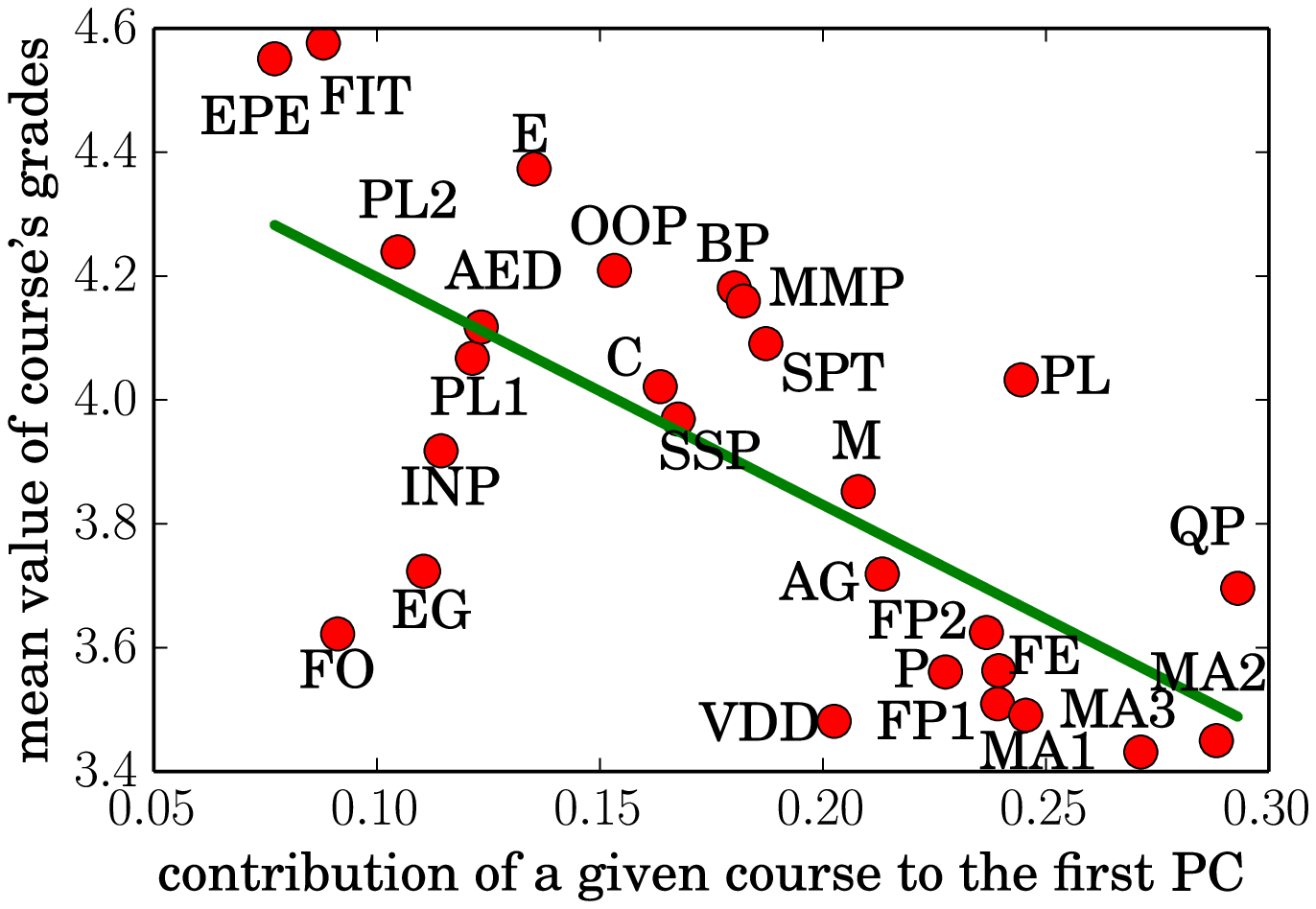}
\caption{$a = -3.7$, $R^2 = 0.5$.}
\label{fig:pcaavs}
\end{subfigure}
\begin{subfigure}[t]{0.5\textwidth}
\includegraphics[height=4.4cm]{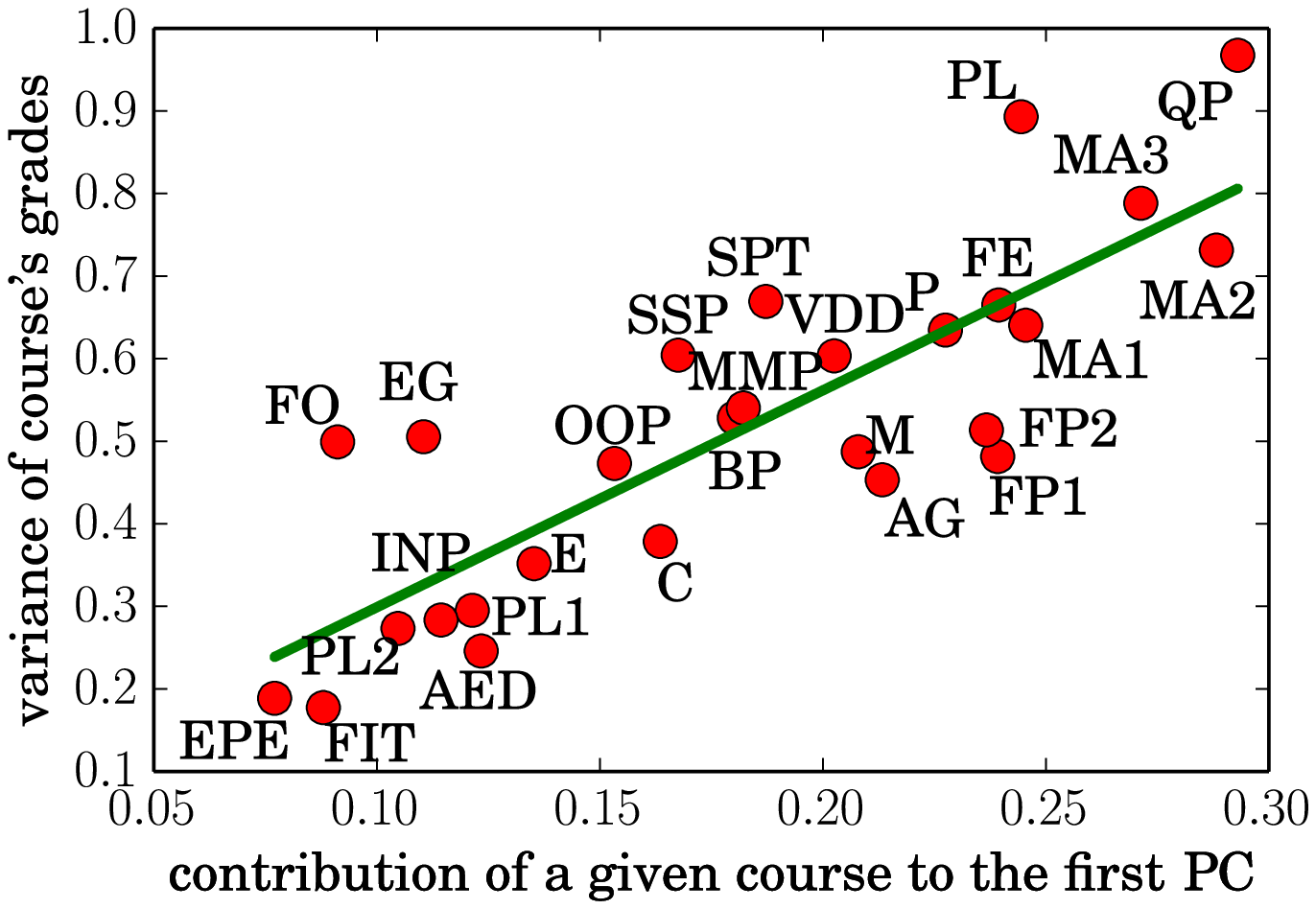}
\caption{$a = 2.6$, $R^2 = 0.7$}
\label{fig:pcavar}
\end{subfigure}
~
\begin{subfigure}[t]{0.45\textwidth}
\includegraphics[height=4.4cm]{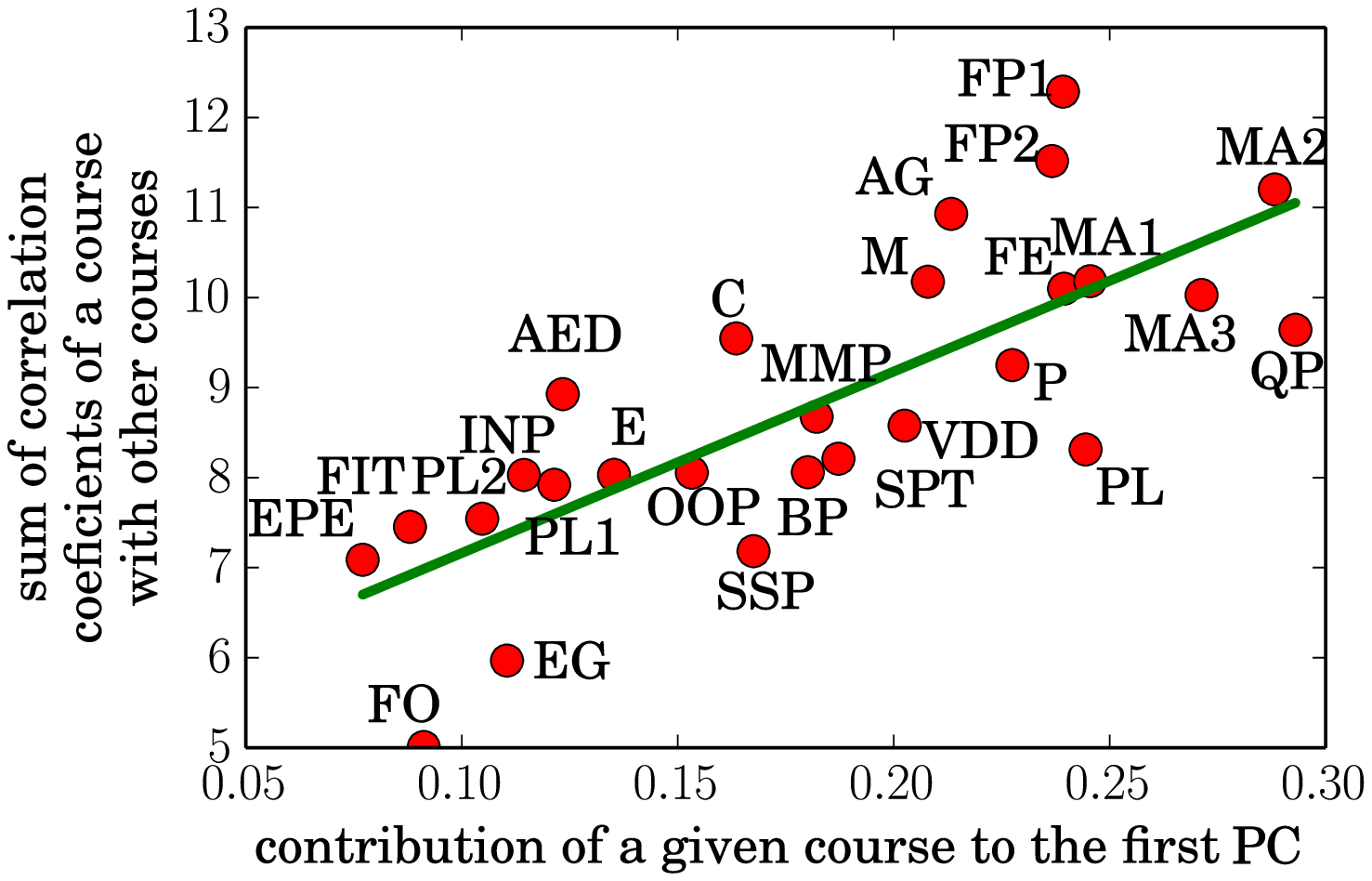}
\caption{$a = 20.15$, $R^2 = 0.6$}
\label{fig:pcacorr}
\end{subfigure}
\caption{(a): variances and means of courses' grades, (b):  courses' mean grades  and  a courses' contribution to the first PC, (c): courses's variances  and  a courses' contribution to first PC, (d):  sum of Pearson's r correlations and courses' contribution to the first PC. Green lines are linear fits ($y=ax+b$), $R^2 $-- coefficient of determination.}
\label{fig:pcavaravs}
\end{figure}

The aim of Fig. \ref{fig:pcavaravs} is to present relations between various measures describing considered courses scores. A slight negative linear correlation ($R^2=0.37$) has been found between mean and variance of course grades (Fig. \ref{fig:avsvar}), i.e. courses with higher mean scores tend to display lower scores variances. Three courses from the group C lie below the trend line at this plot. On the other hand there is group of four courses: QP, MA2, MA3 and PL that possess highest variances and are far above the trend line. We shall call this group \textit{differentiating} (D) since the high variance of scores differences students group. 

A negative linear correlation ($R^2=0.5$) exists also between the course's contribution to 1st PC and the course mean (Fig.\ref{fig:pcaavs}). The largest outlier from the trend is the course FO that was negatively correlated to other courses in Fig. \ref{corr}. Stronger positive correlations are between contributions to 1st PC and variance and sum of correlation coefficients (Figs. \ref{fig:pcavar}, \ref{fig:pcacorr}). In Fig. \ref{fig:pcavar} the C-group of courses lies below and the D-group lies mostly above the trend line. The opposite situation is in Fig. \ref{fig:pcacorr}. Courses from D-group possess the largest contributions to 1st PC and contributions of C-courses are higher than average. 

\section{Concluding remarks}
Statistical analysis combined with data mining methods applied for students' grades make possible to disclose several dependencies between different courses forming academic curriculum at Faculty of Physics, Warsaw University of Technology. A slight negative correlation exists between mean and variance of course grades (Fig. \ref{fig:avsvar}) and between course's contribution to 1st PC and mean scores (Fig.\ref{fig:pcaavs}). Stronger positive correlations are found between the the course's contribution to 1st PC and variance as well as sum of correlation coefficients (Figs. \ref{fig:pcavar}, \ref{fig:pcacorr}). 

The above correlations together with observations of maximal spanning trees (Figs. \ref{fig:mstjkl}, \ref{fig:mstl}) suggest the existence of at least two specific groups of courses in the considered curriculum.

The first group can be called \textit{central} (C) and it consists of two courses {\it Fundamental of Physics I} (FP1), {\it Fundamental of Physics II} (FP2) from semesters 1 and 2, and the course {\it Algebra with Geometry} (AG) from the semester 1. The group C is represented by nodes in centres of maximal spanning trees (Figs. \ref{fig:mstjkl} and \ref{fig:mstl}) since these three courses possess large correlations coefficients with other ones (Table~\ref{tab:corecourses} and Fig. \ref{fig:corrr}). The group C possesses higher than average contributions to 1st PC (Fig. \ref{fig:pcavaravs}) and in Fig. \ref{fig:pcacorr} it is far above the trend line. It means sum of correlations coefficients for these courses is much larger than it can be expected from contributions to 1st PC. However the group C lies below the trend line in Figs.~\ref{fig:avsvar}, \ref{fig:pcavar} i.e. the courses possesses smaller variances of students' scores than it can be expected from their mean grades or their contributions to 1st PC. In other words the courses from the group C reduce differences in initial students' knowledge. One can understand this result as a proof of the proper organization of the considered academic curriculum since lecture topics presented at above courses should make possible a smooth launch of studies.

The second group that can be called \textit{differentiating} (D) consists of courses {\it Quantum Physics} (QP; semester 5), {\it Mathematical Analysis II} (MA2; semester 2), {\it Mathematical Analysis III} (MA3; semester 3) and {\it Programming Languages} (PL; semester 3). This group contributes significantly to students' score variance and to the first principal component (Table \ref{tab:corecourses} and Fig.\ref{fig:pcavaravs}). It follows the above courses highly differentiate the teaching process and their scores separate less skilled students from better ones. This interpretation is confirmed by the fact that above courses possess rather low mean scores (Table \ref{tab:corecourses}, Figs. \ref{fig:avsvar}, \ref{fig:pcaavs}) i.e. corresponding exams are difficult to pass for a large part of students. The group D lies above the trend line in Fig.\ref{fig:avsvar} and below the trend line in Fig. \ref{fig:pcacorr}. The last results mean that this group is less correlated to other courses as one could expect from its contribution to 1st PC. 

Correlations of the first principal component to other courses' parameters seen in Figs. \ref{fig:pcaavs}, \ref{fig:pcavar}, \ref{fig:pcacorr} suggest that the contribution to the first principal component measures a course significance and can be used for assigning ECTS points to a given course. Let us stress that our analyse has been  based entirely on student grades and is independent from declared curricula of considered courses. 

In conclusion, we have shown that data mining methods when applied for students scores can be useful tools for uncovering key courses in academic curricula.  The framework consists of comparing results from the following methods:  (a) observations  of  positions of courses in the Maximal Spanning  Tree corresponding to a correlations matrix, (b) calculations of courses' mean scores and corresponding variances,  (c) calculations of courses' contributions to  the 1st Principal Component.    We suggest that when outcomes from  above methods are similar then such a combined framework    can be used at Universities and at other schools as as an additional tool  to optimize teaching strategies since it does not need  a priori knowledge of assumed curriculum aims. 

\section{Acknowledgements}
This research has received funding as {\it RENOIR} Project from the European Union’s Horizon 2020 research and innovation programme under the Marie Sk\l odowska-Curie grant agreement No 691152. J.A.H. has been also partially supported by Russian Scientific Foundation, proposal \#14-21-0013.

\end{document}